\documentclass[aps,twocolumn]{revtex4}
\usepackage{graphicx}
\usepackage{psfrag}
\usepackage{fancyhdr}

\begin{document}
\date{\today}

\title{Rupture of a liposomal vesicle}

\author {Marco A. Idiart and Yan Levin}

\address{
  Instituto de F\'{\i}sica, UFRGS, 
  Caixa Postal 15051, CEP 91501-970, Porto Alegre, RS, Brazil 
}  

\begin{abstract}
We discuss pore dynamics
in osmotically stressed vesicles. A set of equations which
govern the liposomal size, internal solute concentration, 
and pore diameter is solved numerically.   
We find that dependent on the internal 
solute concentration and vesicle size,  
liposomes can stay pore-free, nucleate a
short lived pore, or nucleate a long-lived pore.  
The phase diagram of pore stability is constructed, 
and the different scaling regimes are deduced analytically.        

\end{abstract}

\maketitle

\section{Introduction}

Liposomal vesicle consists of a lipid bilayer separating
the interior volume, containing an aqueous solution, from
an exterior suspension. The vesicle membrane allows for 
a free exchange of water between the interior and the 
exterior of the liposome, with the flux determined 
by the membrane composition.  On the other hand, 
lipidic membrane strongly inhibits passage of 
large molecules, in 
particular, if they contain ionized groups.  

Liposomes are of great theoretical 
interest as the simplest model of a  biological cell. 
They are also of 
great practical importance as vehicles for 
drug delivery. In the latter case liposomes are designed to
contain a  specific drug or a gene needed to fight the decease.  
The liposomal affinity for infected tissue can be increased by varying
the membrane composition or including ligands which bind
to specific receptors.

If a vesicle containing high internal solute concentration is
placed inside a dilute solution,  the osmotic influx of solvent into
the interior of a vesicle can lead to its rupture.  Whether,
the rupture occurs depends on the membrane elasticity 
and on the internal solute concentration of the liposome. 
Rupture of the liposomal membrane results in  formation of
pores~\cite{TaDvSa75}.  This releases 
the membrane stress,
but comes at a price of exposing the hydrophobic membrane 
interior (lipidic tails) to water.  
Once a pore is formed, the internal content
of the vesicle  begins to leak out, resulting in a decrease of
membrane tension and eventual pore
closure.  We find that depending on the vesicle size
and internal concentration of solute, pores can be either short or
long lived.  For long lived pores a 
scaling relation between the life-time of a pore and the size of
the vesicle is found. The full 
phase diagram of pore stability in the concentration - vesicle size
plane is constructed, 
and the different scaling regimes are deduced analytically.

The paper is organized as follows:  In section \ref{model} we 
review, the previously derived equations 
governing the nucleation and growth of a pore
in an osmotically stressed vesicle~\cite{LeId03}. 
In section \ref{dynamics} numerical
solution of dynamical equations is presented. In section \ref{phase} 
the phase diagram for different dynamical regimes is derived. 
In section \ref{solute} the rate of solute leak-out is determined
and analytical estimates of the pore life-time are provided. 
Finally, in section \ref{conclusions} the conclusions are presented.

\section{ The model}
\label{model}

As was already stressed in the introduction, the liposomal 
membrane allows for a free exchange of water between the
exterior and the interior of a vesicle.  The rate of this
exchange is determined by the permeability of membrane, $P$.  On
the other hand, lipidic membrane strongly inhibits exchange
of solute molecules  between the inside and the outside of
a liposome.   When a vesicle of  high internal 
solute concentration is
placed inside a solute-depleted medium, an osmotic
pressure difference causes an influx of water into
the vesicle.  A vesicle then swells until
the internal Laplace pressure is able to compensate the
the osmotic pressure.  The influx of water 
result in a build up of membrane stress which energetically
favors membrane rupture and formation of pores. Pores are
nucleated in the membrane through thermal fluctuations. Here
we consider the opening of a single pore. The underlying assumption 
is that, once a pore is formed, stress is quickly released
and the creation of a second pore becomes highly unlikely.
This situation is quite different
from what is encountered in electroporation.  In that
case opening of a pore does not fully release the membrane
stress, which is induced by the transmembrane potential,
and one finds a coexistence of pores with 
different sizes\cite{PaChAr79,WeCh96}.

The single pore assumption allow us to write simple 
equations governing the internal vesicle dynamics.
Designating the difference 
between the internal and the external molar 
concentrations of solute as $c$ -- and considering, for mathematical 
simplicity,  a spherical vesicle of radius $R$, and
a circular pore of radius $r$ --- the mass conservation leads to
\begin{equation}
\label{e1} 
4 \pi \rho \; R^2 \;\frac{d R}{dt}= j_w - \pi r^2 \rho v\;,
\end{equation}
where $\rho$ is the density of water,
$j_w$ the osmotic current and $v$ is the leak-out velocity. 

The osmotic current $j_w$ is determined by the 
permeability of the liposomal
membrane and the
difference between the target osmotic pressure  
\begin{equation}
\label{e2a}
\Delta p_o =  k_B T N_A c   
\end{equation}
and the Laplace pressure,
\begin{equation}
\label{e2} 
\Delta p_L=\frac{2 \sigma}{R}\;,
\end{equation}
inside and outside the vesicle. In the above expressions
$k_B$ is the Boltzmann constant, $T$ is temperature, 
$N_A$ is the Avogadro number, and   
$\sigma$ is the membrane surface tension. 
A simple phenomenological expression for the osmotic current
of water into the vesicle is,
\begin{equation}
\label{e3} 
j_w=P (4 \pi R^2-\pi r^2)\left[c - \frac{\Delta p_L }{10^3 k_B T N_A}\right]\;.
\end{equation}
where the conversion factor $10^3$ accounts for the use of molar
concentration of solute $c$.

If $c$ is not too large, the membrane integrity
will not be compromised,  and a stationary state with
$j_w=0$ will be achieved. Under these conditions the osmotic pressure
is completely compensated by the Laplace pressure, 
resulting in a zero net flux
of solvent.  For sufficiently {\it large} internal 
concentration of solute,
a stationary
state will {\it not} be achieved before  membrane  ruptures.  The leak-out 
velocity~\cite{HaBr86,ZhNe93,SaMoBr99,BrGeSa00} of the internal content
of a liposome is  determined by the balance between the
shear stress,  proportional to $\eta v/r$, and the Laplace 
pressure inside the vesicle $\Delta p_L$.  For
low Reynolds numbers~\cite{HaBr86} 
\begin{equation}
\label{e4} 
v=\frac{\Delta p_L \; r}{3 \pi \eta}\;,
\end{equation}
where $\eta$ is the solvent viscosity.

The growth of a pore is controlled by the rate at which the 
membrane elastic energy
is dissipated.  Since the viscosity of membrane is five orders
of magnitude larger than that of water, most of the energy dissipation
is confined to the membrane interior~\cite{DeMaBr95},
\begin{equation}
\label{e5} 
\eta_m l\; \frac{d r}{d t}= \; - \frac{\partial E}{\partial r},
\end{equation}
where $l$ is the membrane width and $\eta_m$ is the membrane viscosity.
A lipid bilayer
has low permeability to solute particles, in particular if they 
are charged, so that the internal solute
concentration
is modified only through the osmotic influx of solvent 
or  the efflux of solute through an open pore, after 
the membrane has ruptured.
The continuity equation expressing this is  
\begin{equation}
\label{e6} 
\frac{4 \pi}{3} R^3 \frac{d c}{d t}=-4 \pi R^2 c \frac{d R}{d t}- 
\pi r^2 c v\;,
\end{equation}
where we have assumed that solute is uniformly distributed inside
the vesicle. In the absence of a pore,  efflux is zero,
and the second term on the right hand side of
Eqs.~(\ref{e1}) and  (\ref{e6}) disappears.

\subsection{The membrane energy}

The membrane energy consists of two terms.  The 
elastic term $E_s$, measuring the cost of increasing the membrane
area beyond its equilibrium unstretched size $A_0$, and the
pore contribution $E_p$ resulting from the partial exposure of   
the hydrophobic lipidic tales to the aqueous environment.

For large osmotic pressures, which are of interest to
us, the membrane thermal undulations can be ignored
and the membrane elastic energy takes a Hooke-like 
form
\begin{equation}
\label{e7}
E_s(R,r) =\frac{1}{2 A_0} \; \kappa \; (A-A_0)^2\;,
\end{equation}
where $A_0=4 \pi R_0^2$ is the equilibrium surface area of an 
unstretched vesicle,  $A =4 \pi R^2 - \pi r^2 $ is the total
membrane area, and $\kappa$
is the membrane elastic modulus~\cite{HeSe84,EvRa90,SeIs02}.
The membrane surface tension is
\begin{equation}
\label{e8a}
\sigma = \frac{\partial E_s}{\partial A} =  \kappa \; \frac{A - A_0}{A_0} \;,
\end{equation}
and the pore energy is 
\begin{equation}
\label{e8}
E_p(r) = \; 2 \; \pi \; \gamma \; r \;,
\end{equation}
where $\gamma$ is the pore line tension. 

The typical values for
the physical constants involved in the model are given
in the Table 1.
\begin{table}
\begin{center}
\begin{tabular}{lll} \hline \hline
Parameter & Value & Source  \\ \hline
 $\gamma$ & $ 10^{-12} \; J/m $ & Ref. \cite{BrGeSa00} \\
 $\kappa$ & $0.2  \; J/m^2 $ & Ref. \cite{EvRa90} \\
 $\eta_m$ &  $ 100 \; Pas $ & Ref. \cite{BrGeSa00} \\
 $\eta_w$ & $ 0.001 \; Pas$ & - \\
 P & $1.8 \times 10^{-4} \; kg/(m^2 s M ) $ & Ref. \cite{Sa83} \\
 $l$ & $3.5 \; nm$ & Ref. \cite{BrGeSa00} \\
\hline
\end{tabular}
\end{center}
\caption{Characteristic values for the physical parameteres used in the calculations.}
\end{table}

\subsection{The rupture condition}

\begin{figure}[h]
\begin{center}
\includegraphics[angle=270,width=6cm]{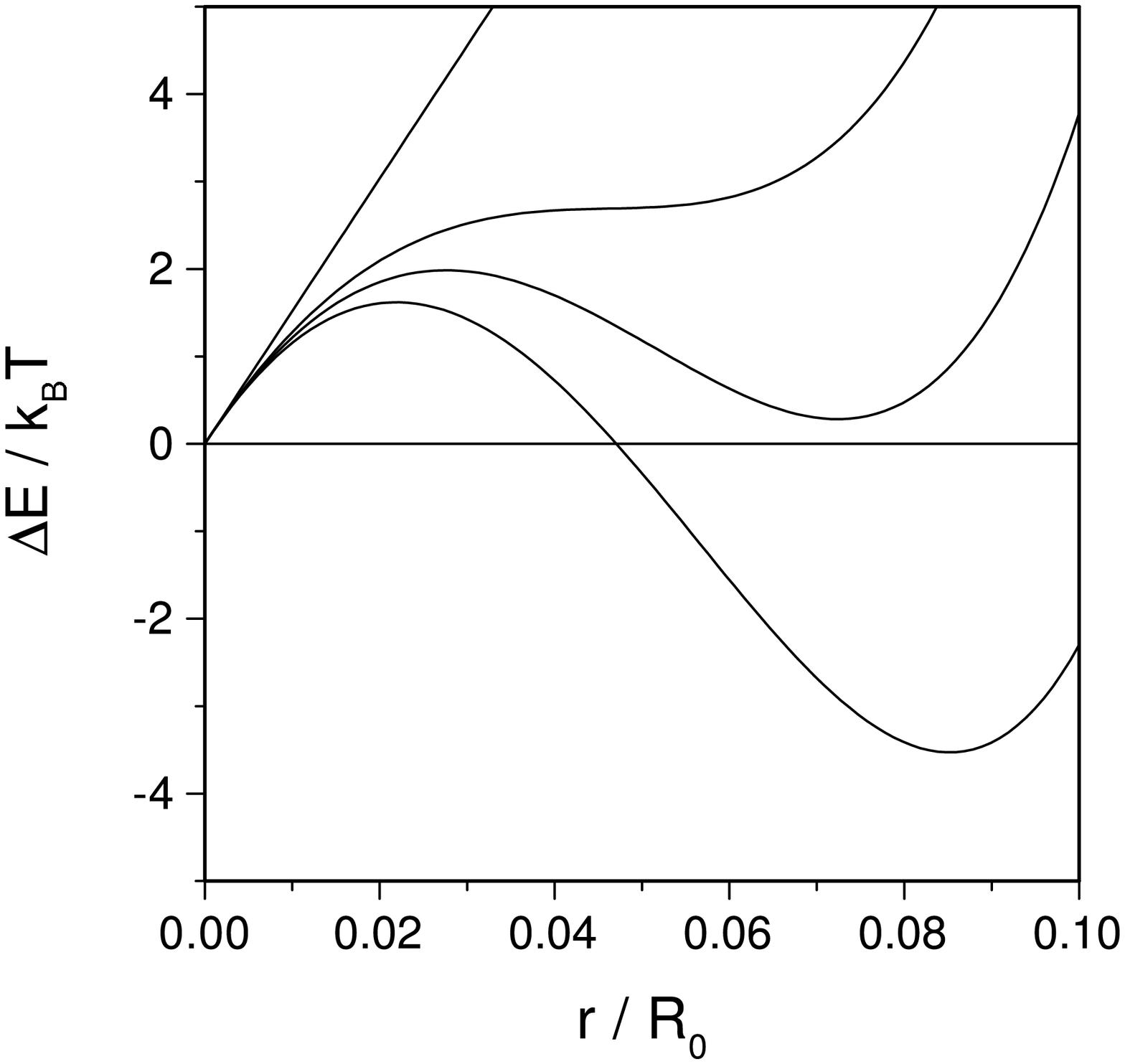}
\end{center}
\caption{Energy $\Delta E(R,r)$ necessary to open a pore of 
radius $r$ in a liposome with $R_0=100 \,nm$ and ratio
$\gamma/(\kappa \; R_0) = 5\times 10^{-5}$. 
The curves are, from top down,  $R/R_0=1.0$, $R/R_0=1.0008$,
$R/R_0=1.001$, and $R/R_0=1.0012$ }
\label{fig1}
\end{figure}

The total energy of a membrane containing a pore is  
$E(R,r)=E_s(R,r)+E_p(r)$. A cost of opening a pore
of radius $r$ is then,
\begin{equation}
\label{e8b}
\Delta E(R,r)=E(R,r)-E(R,0) \;.
\end{equation}
In Fig. \ref{fig1} we plot $\Delta E(R,r)$ as a function of $r$
for various ratios of 
$R/R_0$. For $R \simeq R_0$, the membrane is relaxed and  $r=0$ is the only 
minimum of $\Delta E$.  
For $R$ bigger than the critical radius $R_c$, the energy cost
function develops a barrier located at  
\begin{equation}
\label{e9}
r_b = \frac{4 \sqrt{R^2 - R_0^2}}{\sqrt 3} \; 
\cos\left( \frac{\varphi-2\pi}{3} \right) \;
\end{equation}
and a new minimum at 
\begin{equation}
\label{e10}
r_m = \frac{4 \sqrt{R^2 - R_0^2}}{\sqrt 3} \; \cos\left( \frac{\varphi}{3} \right) \;,
\end{equation}
where
\begin{equation}
\label{e11}
\varphi(R,R_0) = \cos^{-1} \left( - \frac{3\sqrt3}{8} 
\frac{E(R, 2 \sqrt{R^2 - R_0^2})}{E(R,0)} \right) \;.
\end{equation}
The critical vesicle size $R_c$ for appearance of a new minimum 
is determined by the condition $\varphi=\pi$ or 
\begin{equation}
\label{e11a}
\frac{3\sqrt3}{8} \frac{E(R_c,2 \sqrt{R_c^2 - R_0^2})}{E(R_c,0)} = 1 \;.
\end{equation}
Substituting the solution of Eq.~(\ref{e11a}) into Eq.~(\ref{e8a})
the critical surface tension for appearance of the second minimum is 
\begin{equation}
\label{e12}
\sigma^{(1)}_c = 3 \left( \frac{\gamma \; \sqrt{\kappa}}{R_0} \right)^{2/3} \;.
\end{equation}
>From Eq. (\ref{e3}) we see that a minimum solute   
concentration 
\begin{equation}
\label{e13}
c^{(1)}_{min} = \frac{ 2 \; \sigma^{(1)}_c }{ 10^3 k_B T N_A R_0 
\sqrt{1+\sigma^{(1)}_c/\kappa} } \;,
\end{equation}
is necessary to develop the second minimum in $\Delta E$
at $r_m$.  
However, concentration 
$c^{(1)}_{min}$ does not guarantee opening of a pore.  
Even if $\Delta E(R,r_m)< 0$, the energy barrier to pore nucleation
can be many $k_B T$ high. Therefore, pores with radius 
less than $r_b$ will 
quickly re-seal, without having a chance to grow.  
The probability of occurrence of a sufficiently large thermal 
fluctuation necessary to open a pore with  $r>r_b$ is
\begin{equation}
\label{e14} 
P(r) \sim e^{-\beta \Delta E(R,r_b)}\;.
\end{equation}
The waiting time for opening a 
pore of radius $r \ge r_b$ is, therefore, very long 
unless 
\begin{equation}
\label{e15}
\Delta E(R_p,r_b)\simeq k_B T  \;.
\end{equation}
This equation, then,  determines the size of a swollen
vesicle
\begin{equation}
\label{e15a}
R_p=R_0\sqrt{1+\sigma^{(2)}_c/\kappa} 
\end{equation}
which is able to nucleate a growing pore.  
The membrane tension of such a liposome is approximately,
\begin{equation}
\label{e17}
\sigma_c^{(2)} \; \approx  \; \frac{\pi \gamma^2}{k_B T}
\end{equation}
and the critical pore size is $r_b \approx \gamma/\sigma$.
The minimum concentration of solute necessary to reach this
tension is
\begin{equation}
\label{e13a}
c^{(2)}_{min} = \frac{ 2 \; \sigma^{(2)}_c }{ 10^3 k_B T N_A R_p } \;.
\end{equation}

We note that for membrane parameters given in Table 1,
the membrane tension 
is $\sigma^{(2)}_c \simeq 10^{-3}\, J/m^2$, 
which is very close to the one
found to be necessary to rupture a mechanically 
stretched membrane~\cite{Ti74,WeCh96}.

\section{Pore dynamics}
\label{dynamics}

\begin{figure}[h]
\vspace{0.5cm}
\begin{center}
\includegraphics[angle=270,width=8cm]{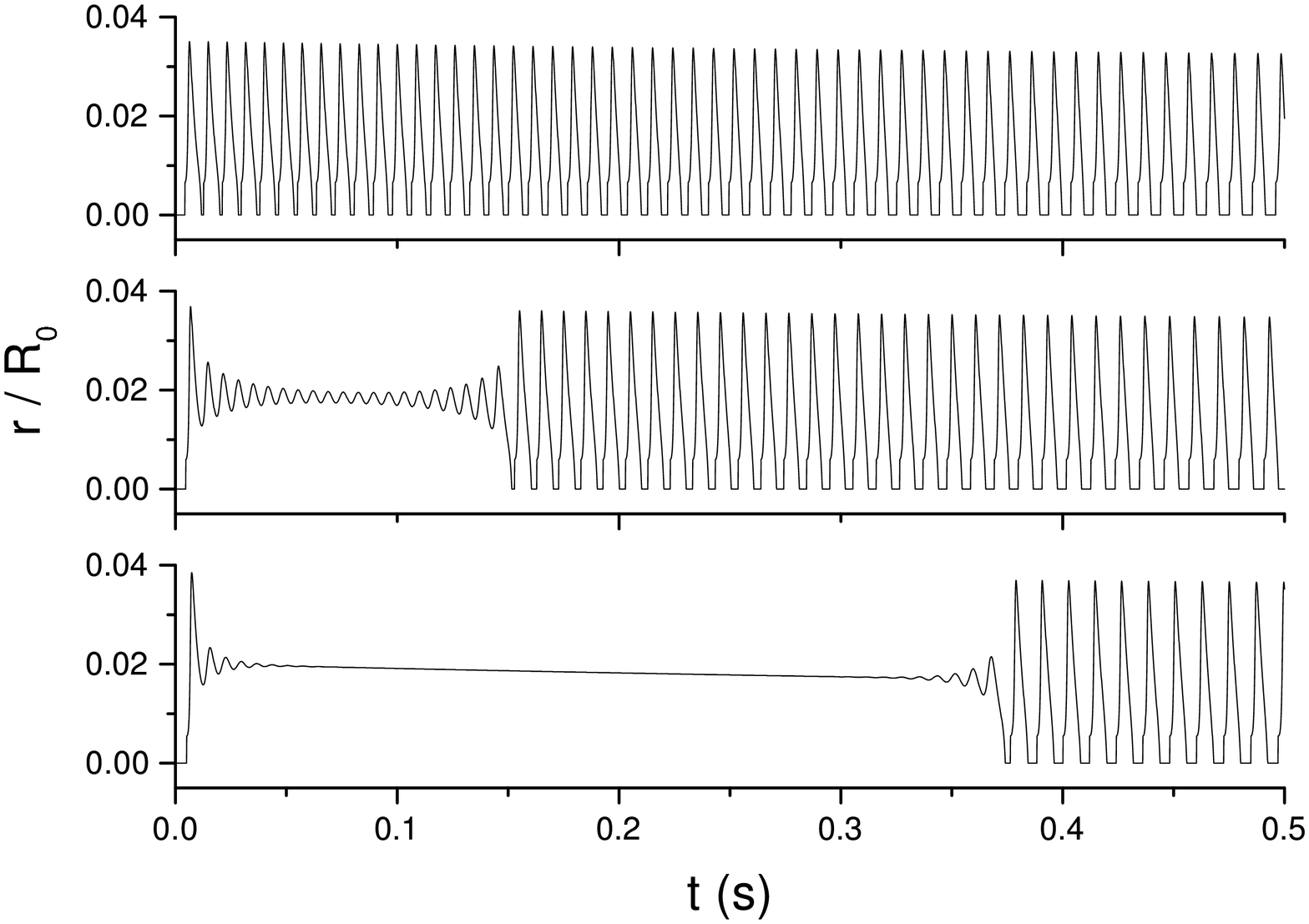}
\end{center}
\vspace{-0.5cm}
\caption{Radius of a pore as a function of time for vesicles
of $R_0=200\, nm$ (topmost), $220 \,nm$ and $240\, nm$ (bottommost), for
initial concentration $c_0=0.5 M$.
Note that for vesicle with $R_0=200 \,nm$ the pores are short-lived,
while for larger vesicles, long-lived pore opens first.}
\label{fig2}
\end{figure}

The rupture dynamics of an osmotically stressed vesicle proceeds as follows.
At $t=0$ the vesicle starts swelling, its size and internal concentration
controlled by Eqs.~(\ref{e1}) and (\ref{e6}).  As it swells,
the membrane surface tension increase until $\sigma=\sigma^{(1)}_c$ 
and the energy function develops a new minimum. If the
barrier height is less then $k_B T$, a pore of size $r_b$
is nucleated. From this moment the dynamics of the vesicle evolution is 
controlled by the set of  Eqs.~(\ref{e1}), (\ref{e5}) and (\ref{e6}) .  
On the other hand, if $\Delta E(R_c,r_b)>k_B T$, the  
swelling continues 
without a pore nucleation until
Eq.~(\ref{e15}) is satisfied and a pore of
radius $r_b$ opens.  
After a pore is nucleated,
the internal content of the vesicle begins to leak-out, decreasing
the membrane tension and leading to an eventual re-sealment of the pore.
The cycle will  be repeated until the internal concentration
of solute drops bellow $c^{(1)}_{min}$ and a 
steady state with $j_w=0$ is established. 
In Fig. \ref{fig2} we show the pore radius as a function of time
for vesicles of 
three different sizes and initial concentration of solute $c_0=0.5 M$.   
We see that small vesicles are characterized by 
rapid opening and closing of 
pores, resulting in a  periodic flickering with a
characteristic time $\tau_f \approx 10^{-2} s$.  
On the other hand, larger vesicles are capable of nucleating 
a long-lived pore.
\begin{figure}[h]
\begin{center}
\vspace{-0.5cm}
\includegraphics[angle=270,width=10cm]{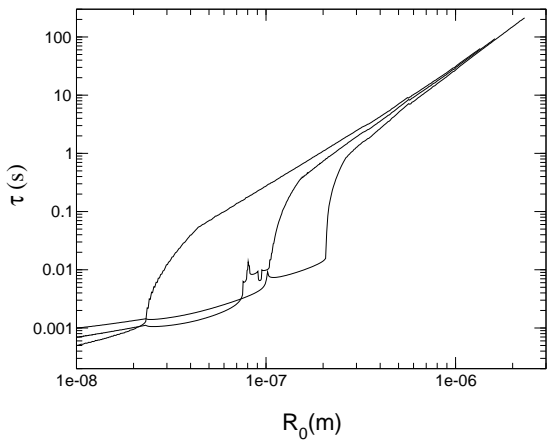}
\end{center}
\vspace{-1cm}
\caption{Life span of the first-open pore as a function of 
vesicle size $R_0$ for $c_0=0.5 M$ (rightmost), 1.0M and 5.0M(leftmost).  
Note the appearance of critical vesicle
size $R_0^c(c_0)$ (sharp change in slope of $\tau$ vs. $R_0$) 
which sustains a long-lived pore. The peculiar
spikes  in the life-time of vesicles containing low concentration
of solute, are an artifact of the way the pores are nucleated. On the other
hand existence of $R_0^c(c_0)$ is independent of pore nucleation protocol.}
\label{fig3}
\end{figure}
After the long-lived pore has closed, it is followed by a sequence of
short lived pores, with the characteristic life span $\tau_f$. 
The life-span of a long-lived pore $\tau$ is show in Fig. \ref{fig3}.
For large vesicles, life-span scales with the vesicle size as
\begin{equation}
\label{e18} 
\tau \sim R_0^\nu\;,
\end{equation}
with $\nu \approx 2.3-2.4$, see Fig.\ref{fig3}.

\section{Phase diagram}
\label{phase}

\begin{figure}[h]
\begin{center}
\includegraphics[angle=-90,width=8.8cm]{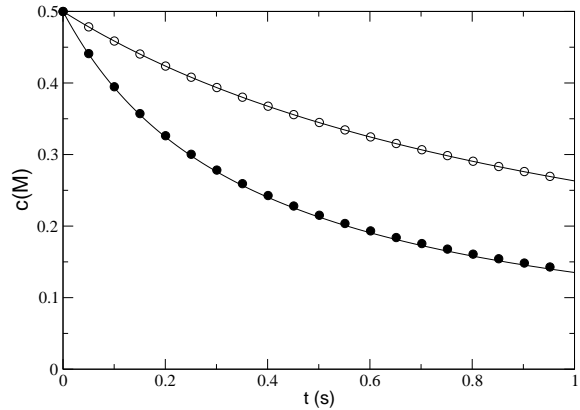}
\end{center}
\vspace{-1cm}
\caption{Concentration decay as a function of time.
The circles are the result of numerical integration of Eqs.~(\ref{e1}),
(\ref{e5}) and (\ref{e6}). The solid lines are 
from the analytical expression (\ref{e23k}).
Open circles are for vesicle of $R_0 = 100 nm$ 
and filled circles  are for vesicle of $R_0=300 nm$. The initial
solute concentration is $c_0=0.5 M$. }
\label{fig4}
\end{figure}

To better understand the details of the vesicle evolution,
it is convenient to separate the membrane
dynamics from the concentration dynamics. Since
the internal solute concentration changes very slowly compared
to the $\tau_f$, see Fig. \ref{fig4}, 
as a first approximation we can take it to be 
constant.  

\begin{figure}[h]
\vspace{0.5cm}
\includegraphics[angle=0,width=7.5cm]{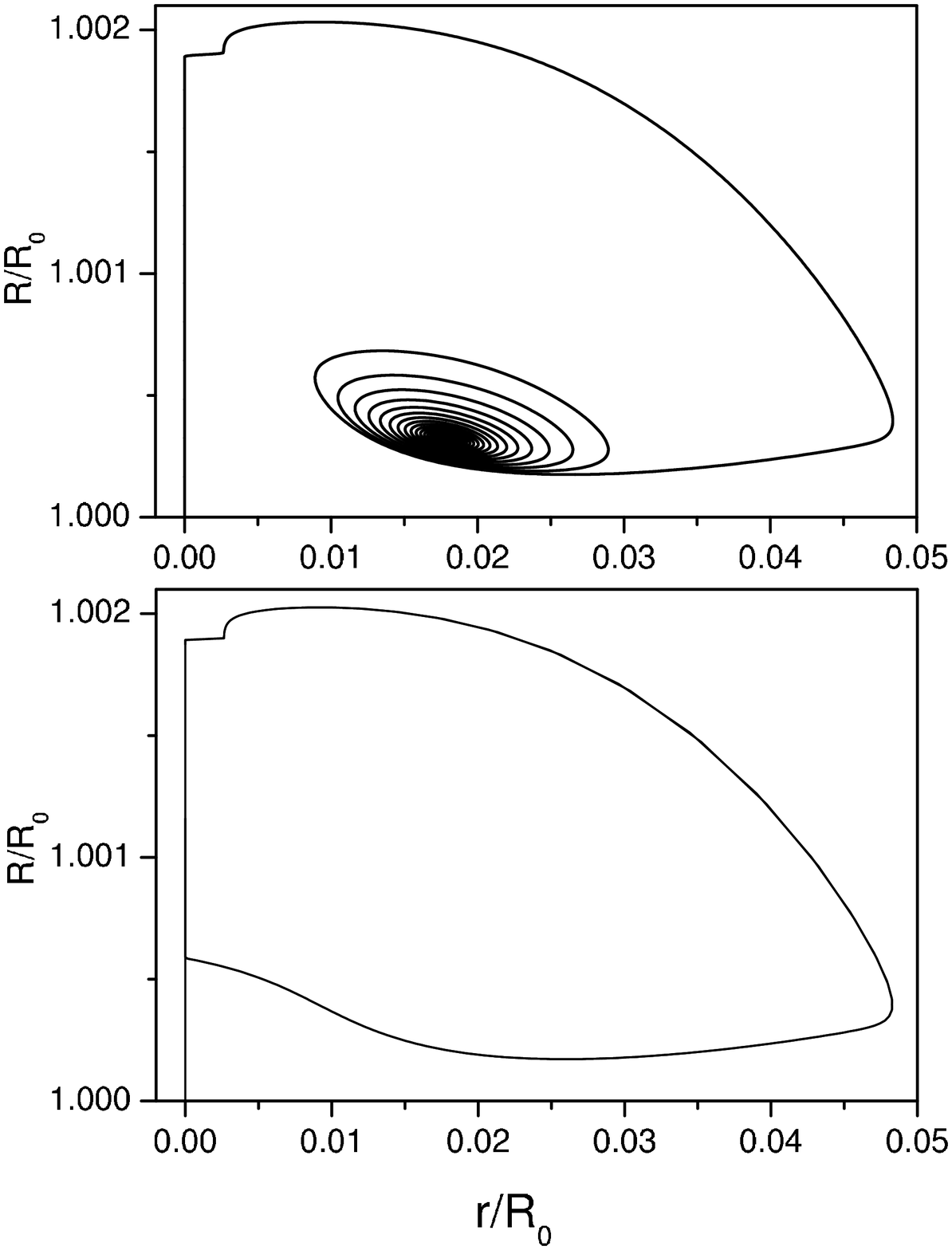}
\vspace{-1cm}
\caption{Dynamics of the pore for fixed concentration
and $R_0=500 nm$. The top panel is for $c_0=0.19 M$, and 
the bottom panel for $c_0 = 0.18 M$.}
\label{fig5}
\end{figure}

\begin{figure}[h]
\vspace{0.5cm}
\hspace{-0.5cm}
\includegraphics[angle=0,width=9cm]{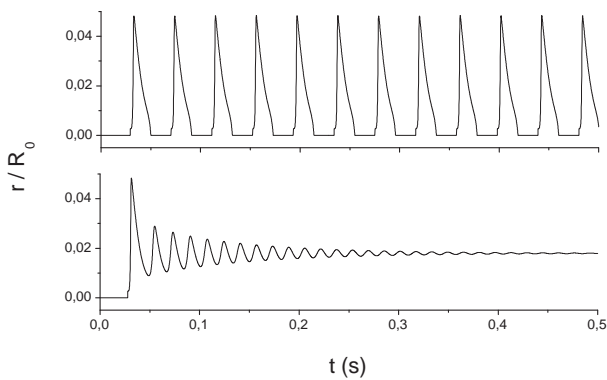}
\vspace{-1cm}
\caption{
Radius of a pore as a function of time for vesicles
of $R_0=500\, nm$ for a constant internal concentration
of solute $c_0=0.18M$ (top) and $c_0=0.19M$(bottom).}
\label{fig6}
\end{figure}

\begin{figure}[h]
\vspace{0.5cm}
\hspace{-1.0cm}
\includegraphics[angle=-90,width=9cm]{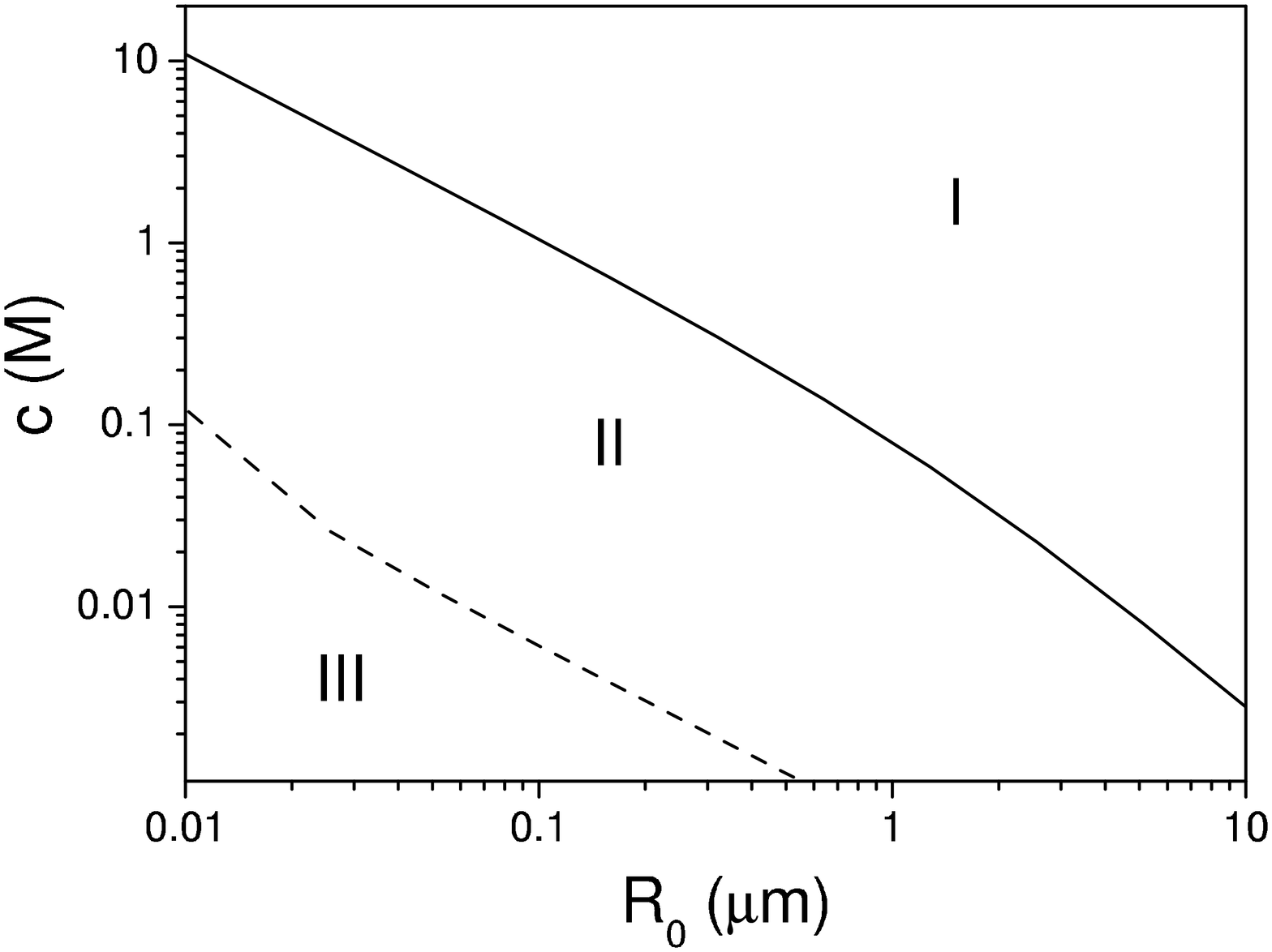}
\caption{The phase diagram for the constant 
concentration dynamics. In the region 
I, an open pore is stable. In region II, size of a pore
oscillates, and in region III  pored do not open. The 
dashed line indicates the minimum solute concentration  
needed to open a pore. The discontinuity in
its derivative is due to the fact that for small $R_0$ 
the barrier to pore nucleation is lower 
than $k_B T$. 
}
\label{fig7}
\end{figure}

In this case the vesicle evolution
is controlled by  Eqs.~(\ref{e1}) and (\ref{e5}), which
can be written as 
\begin{equation}
\label{e19}
\frac{dR}{dt} = F(R,r) 
\end{equation}
\begin{equation}
\label{e20}
\frac{dr}{dt} = G(R,r)
\end{equation}
where
\begin{equation}
\label{e21}
F(R,r) = \frac{1}{4 \pi \rho R} ( j_w - \pi r^2 \rho v )
\end{equation}
and
\begin{equation}
\label{e22}
G(R,r) = - \frac{1}{\eta_m \ell} \frac{\partial E}{\partial r} \;.
\end{equation}
The vesicle dynamics 
is governed by the fixed point ($r^*$,$R^*$), determined from
$dr/dt=0=G(R^*,r^*)=0$ and $dR/dt=0=F(R^*,r^*)$. The stability
of the fixed point is controlled by the eigenvalues 
$\lambda_1$ and $\lambda_2$  of the Jacobian
matrix
\begin{equation}
\label{e22a}
J=\pmatrix{\frac{\partial F}{\partial R}&\frac{\partial F}{\partial r}\cr\frac{\partial G}{\partial R}&\frac{\partial G}{\partial r} } \;.
\end{equation}
It is important to keep in mind that the coefficients of the Jacobian
matrix are real and,  therefore, the eigenvalues are either real or  complex conjugates, 
$\lambda_1=\bar \lambda_2$.  For all the  parameters that we have investigated 
the eigenvalues are complex conjugates, and the stability is
governed by  $Re(\lambda_1)=Re(\lambda_2)\equiv Re(\lambda)$.
If $Re(\lambda)<0$, the fixed point is stable  
and a stationary state with a pore of size $r^*$ and 
vesicle of size $R^*$ will
be established. On the other hand if  $Re(\lambda) \ge 0$ the
fixed point is unstable, and the pore will eventually close, 
see Figs. \ref{fig5} and \ref{fig6}. 
A new pore will open when the membrane tension again reaches 
the  value $\sigma^{(2)}_c$.  This process will repeat indefinitely
with the characteristic time $\tau_f$.
The phase boundary, in the concentration-size plane $(R_0,c)$,
between the two dynamical regimes is determined by the 
condition $Re(\lambda)=0$ or equivalently
\begin{equation}
\label{e23}
\left. Tr J \; \right|_{R^*,r^*}= 0 \;, 
\end{equation}
which reduces to
\begin{equation}
\label{e23a} 
\left. \frac{1}{4\pi \rho R^2} \frac{\partial}{\partial R}(  j_w  - \pi r^2 \rho  v )
 - \frac{1}{\eta_m \ell}\frac{\partial^2 E}{\partial r^2} \right|_{R^*,r^*} = 0 \;.
\end{equation} 
Combining Eqs. (\ref{e2}) and (\ref{e4}), 
the leak-out velocity is 
\begin{equation}
\label{e23b} 
v=\frac{2 \sigma  r}{3 \pi \eta R} \;. 
\end{equation}
Differentiating $v$ with respect to the vesicle size  
\begin{equation}
\label{e23c} 
\frac{\partial v}{\partial R} = - \frac{v}{R}+ 
\frac{16 r \kappa}{3 \eta A_0}\;.
\end{equation}
The change of surface tension with the pore size is given by
\begin{equation}
\label{e23d} 
\frac{\partial \sigma}{\partial r} = - 2\pi r \frac{\kappa}{A_0} \;.
\end{equation}
For large concentrations, and $R \gg r$, 
the osmotic current can be approximated by
\begin{equation}
\label{e23e}
j_w  \simeq P A \; c \;,
\end{equation}
so that
\begin{equation}
\label{e23f}
\frac{\partial j_w}{\partial R} = \frac{2}{R} j_w
\end{equation}
At fixed point $(R^*,r^*)$,
\begin{equation}
\label{e23g}
j_w^*=\pi r^{*2} \rho v^*=\frac{2 \rho \gamma^3}{3 \eta R_0 \; \sigma^{*2}}\;,
\end{equation}
where we have approximated $R \simeq R_0$ and 
$r^*=r_b \simeq \gamma/\sigma^*$. 
Eq. (\ref{e23a}) then reduces to a quartic equation for $\sigma^*$
\begin{equation}
\label{e23i}
\sigma^{*4} + \frac{2\gamma^{2}}{A_0} \left( 
\frac{\eta_m \ell \gamma }{\eta R_0^2} - \pi \kappa \right) \sigma^* - 
\frac{16 \pi }{3} \frac{\eta_m \ell \gamma^3 \kappa}{\eta A_0^2} = 0 \;.
\end{equation}
Combining Eqs. 
(\ref{e23e}) and (\ref{e23g}), the phase boundary 
separating the regime of short-lived pores (region II) from the 
long-lived (infinite life-time) pores (region I) is 
\begin{equation}
\label{e23h}
c_c(R_0)=\frac{ \rho \gamma^3}{6  \pi \eta \; P  R_0^3 \sigma^*}\;,
\end{equation}
see,  Fig.\ref{fig7}.

For small concentrations, the
surface tension does not build sufficiently 
high to cause the membrane 
rupture (region III). The phase boundary
between regions II and III is denoted by a dashed curve in Fig.\ref{fig7}.
The discontinuity in slope results from the nucleation barrier 
passing the threshold $\Delta E =k_B T$.  Thus the right hand side of
the II-III phase boundary is given by $c_{min}^{(2)}(R_0)$, while the 
left hand side $c_{min}^{(1)}(R_0)$

The role of the concentration dynamics is to make the system traverse 
through the different regions of the phase diagram, 
controlling the time of
permanence in each regime.  


\section{Dynamics of solute leak-out}
\label{solute}

The critical size of a liposome $R_0^c(c_0)$, necessary for
nucleating a long-lived pore, depends on the
initial solute concentration. 
The larger is the  solute concentration $c_0$, 
the smaller will be the size of a vesicle which supports a
long-lived pore.   The long life-span 
of these pores is the result of a ``wash-out'' effect in which
the osmotic flux is almost completely compensated by the leak-out rate 
of solute through the pore. 
When solute
concentration inside the vesicle drops below the critical value,
$c_c(R_0)$,  the long-lived pore closes. This value is 
insensitive to  the initial solute concentration $c_0$, 
but depends strongly on the 
vesicle size $R_0$.

Combining Eqs.(\ref{e1}) and (\ref{e6}) we obtain
\begin{equation}
\label{e23j}
\frac{4 \pi}{3} R^3 \frac{dc}{dt} \; = \; -  \;
\frac{j_w}{\rho} \; c \;.
\end{equation}
Approximating $j_w \simeq P c A$  
and $R\simeq R_0$  leads to
\begin{equation}
\label{e23k}
c(t) = c_0 \left[ \frac{ 3 P c_0}{\rho R_0} \; t \; + \; 1 \;
\right]^{-1}
\end{equation}

Eq. (\ref{e23k}) provides an almost  perfect
fit of the time dependence of the internal solute  
concentration, see
Fig.\ref{fig4}. The life span of a long lived pore can be approximated by
the time it takes for the solute to go from the initial
concentration $c_0$ to the critical concentration
$c_c(R_0)$, below which the pore is no longer stable,
\begin{equation}
\label{e23l}
\tau \; \simeq \; \frac{\rho R_0}{3 P c_c(R_0)} \left[ 1 - \frac{c_c(R_0)}{c_0} \right]
\end{equation}

It is possible to derivate two limits
for the critical concentration $c_c(R_0)$, see Appendix.
Writing
\begin{equation}
\label{e24}
c_c(R_0) = \frac{\rho \gamma}{ 6\pi \eta P} \frac{1}{R_0 f(R_0)}
\end{equation}
for vesicles of radius $R_0 \simeq R_1 \equiv \sqrt{\frac{\gamma \eta_m \ell}{ \pi \kappa \eta}}$
\begin{equation}
f(R_0) = \left[ \frac{1}{3^{1/4}} \sqrt{ \frac{\kappa}{\gamma} R_1 }
 + \frac{\sqrt{3}}{8 R_0 R_1 } \left( R_0^2 - R_1^2 \right) \right]^2
\label{e24a}
\end{equation} 
and for $R_0 >>  R_1$
\begin{equation}
f(R_0) = 
\left[\frac{1}{2} \frac{\kappa}{\gamma} \frac{1}{R_0} \left(  {R_0}^2 - {R_1}^2  
   \right) \right]^{\frac{2}{3}}
\label{e24b}
\end{equation}
and therefore the life-span of long-lived pores scales as
\begin{eqnarray}
\tau \sim \left\{ \begin{array}{ll}  R_0^{2}, & R_0 \simeq R_1 \\
                               R_0^{2+2/3}, & R_0 >> R_1 \;.
            \end{array} \right. 
\end{eqnarray}
For the parameters used in this paper $R_1= 41.8 nm$,
so that Eq.~(\ref{e24b}) is consistent with the numerical findings,
see Fig. \ref{fig3}.

The flickering time $\tau_f$ is approximately the time it takes for a
vesicle to swell to size $ R_p $, needed to induce 
a liposomal rupture.  During the swelling, 
internal concentration of solute changes very
little, since $R_p \simeq R_0$, so that $c$ can be kept
constant.  Furthermore, for large
initial solute concentrations, the osmotic current is 
$j_w \approx 4\pi P R^2 c$, and Eq.~(\ref{e1}) is easily
integrated yielding,
\begin{equation}
\label{e25}
\tau_f=\frac{\rho (R_p-R_0)}{P c} \;.
\end{equation}
On the other hand
\begin{equation}
\label{e26}
\sigma_c^{(2)}=\kappa \frac{R_p^2-R_0^2}{R_0^2}\approx 
\frac{2 \kappa}{R_0} \left(R_p-R_0\right) \;.
\end{equation}
Substituting Eq.~(\ref{e26}) into Eq.~(\ref{e25}) we obtain
the expression for the flickering time,
\begin{equation}
\tau_f \approx \frac{\rho \; R_0 \; \sigma_c^{(2)}}{ 2 \; P \; c \; \kappa} \;,
\end{equation}
where $\sigma_c^{(2)}$ is given by Eq.~(\ref{e17}).


\section{Conclusions}
\label{conclusions}
 
We have presented a theory for nucleation and growth of pores in
osmotically stressed liposomal vesicles.  The model predicts
that depending on the internal solute concentration and the liposome
size, pores can be either short-lived --- opening and closing with
a characteristic time $\tau_f$ --- or long-lived, with their 
life-time scaling with the size of the vesicle. 
   
Long lived pores have been observed in
red blood cell ghosts~\cite{St70,StKa74}. No theory,
up to date, was able to account for these long-lived pores.
Holes were predicted to either grow indefinitely,
which would result in ghost vesiculation, or to 
close completely~\cite{BeBr99}.  
Our model 
provides a dynamical mechanism for pore stabilization, consistent
with the experimental observations.  However, for the specific 
case of red blood cell ghosts the ratio of $\gamma/\kappa$ must be
adjusted to obtain the pore size observed in experiments.  This
is not surprising since the real biological cells, unlike liposomes,
have a complicated internal cytoskeleton, which  strongly affects
the membrane elasticity. 

In aqueous solutions the phospholipid membranes
acquire a net negative charge. At physiological
concentrations, $154\,mM$ of $NaCl$, the Debye length, however, is
quite short, less then $1\,nm$ 
and the electrostatic interactions are strongly
screened~\cite{Le02}.  We, therefore, do not expect that electrostatics 
will significantly modify the basic conclusions  of our
theory, beyond the renormalization of membrane line~\cite{BeBr99} 
and surface tension.  However, further, investigations in this direction
are necessary and will be the subject of future work.

Finally, up to now we have not taken into account a diffusive
efflux of solute through an open pore.  The characteristic
time for effusion can be estimated as~\cite{LeIdAr03} 

\begin{equation}
\label{15} 
\tau_e \approx \frac{R_0^3}{ r D}\;,
\end{equation}
where $D$ is the diffusion constant.
Using $D \approx 10^{-9}\, m^2/s$, appropriate for small 
organic molecules such
as sucrose,
and $r=r^*\approx \gamma/\sigma_c \approx 1 \,nm$, we see that for liposomes 
with $R_0=200 \,nm$, the time for effusion is 
$\tau_e \approx 10^{-2}$ s.  This is comparable to
the flicker time $\tau_f$. Therefore, 
for small vesicles effusion is 
an important mechanism for loss of solute.  On
the other hand, for large liposomes with $R_0= 500 \,nm$ and 
above, effusion is only marginally relevant.

This work was supported in part by the Brazilian agencies
CNPq and FAPERGS.    

\section{Appendix}

Here we present a derivation of the limiting form of the phase boundary, 
Eqs.~(\ref{e24})-(\ref{e24b}),
separating the region I and II of the phase diagram, Fig. \ref{fig7}. 
Writing 
\begin{equation}
 \sigma^* = \frac{\gamma}{R_0} \; x \;,
\label{a1c} 
\end{equation}
Eq. (\ref{e23i}) reduces to 
\begin{equation}
x^4 + \Gamma_1 x + \Gamma_2 = 0
\label{a2}
\end{equation}
with 
\begin{eqnarray}
\Gamma_1 &=& - \frac{1}{2}  \frac{ \kappa }{\gamma }\left( 1 + \frac{ R_1}{R_0} \right) ( R_0 - R_1 ) \\
\Gamma_2 &=& - \frac{1}{3} \left( \frac{\kappa}{\gamma} R_1 \right)^2
\end{eqnarray}
where
$$
R_1 = \sqrt{\frac{ \gamma \; \eta_m \ell}{ \pi \kappa \; \eta}} \;.
$$

The solutions of a quartic equation like (\ref{a2}) can be written as
\begin{eqnarray}
x_{1,2} = \frac{1}{2} \left[ \;\;\; \sqrt y  \pm \sqrt{ - | y | - \frac{2\Gamma_1}{\sqrt y} } \;\; \right] \nonumber \\
x_{3,4} = \frac{1}{2} \left[ \; - \sqrt y  \pm \sqrt{ - | y | + \frac{2\Gamma_1}{\sqrt y} } \;\; \right] \nonumber 
\end{eqnarray}
where y is the real root of the resolvent
$$
y^3 - 4\Gamma_2 \; y - \Gamma_1^2 = 0 \;.
$$
Since the surface tension is non-negative, 
the physically relevant solution 
for x  depends on the sign 
of $\Gamma_1$ and  we can write 
\begin{eqnarray}
x = 
\frac{1}{2} \left[ \sqrt{\frac{2|\Gamma_1|}{\sqrt y}-|y|} - \mbox{sgn}(\Gamma_1)
\sqrt y  \;\; \right]  \;.
\end{eqnarray}
For an equation of the form $y^3 + a y + b = 0$  there is a single real root if
$$
Q = \left(\frac{b}{2}\right)^2 + \left(\frac{a}{3}\right)^3 > 0
$$ 
and this real root is
\begin{equation}
\label{y}
y = \left( -\frac{b}{2} + \sqrt Q \right)^{1/3} + 
\left( -\frac{b}{2} - \sqrt Q \right)^{1/3}
\end{equation}
This is precisely our case since $\Gamma_2 < 0$ and, therefore, $Q>0$. 
It is convenient to rewrite Eq.{\ref{y}} as
$$
y =  \left( \frac{ \Gamma_1^2}{2} \right)^{1/3} \; \psi( z )
$$
where
$$
z = - \left( \frac{ \Gamma_2}{3} \right)^{3} \; \left( \frac{4}{\Gamma_1} \right)^{4}
$$
and
$$
\psi( z ) =
\left( \sqrt{1+z} + 1  \right)^{1/3} -
\left( \sqrt{1+z}-1 \right)^{1/3}  \;.
$$
The psi function has the following asymptotic behaviors 
\begin{eqnarray}
\psi(z) = \left\{ \begin{array}{ll}  \frac{2}{3} z^{-1/3} - \frac{8}{81} z^{-4/3}  & z  >> 1 \\ & \\
                                           2^{1/3} - \left( z/2 \right)^{1/3}            & z  << 1  \;.
                       \end{array} \right.
\end{eqnarray}
Putting everything together we have
\begin{eqnarray}
y = \left\{ \begin{array}{ll}  \frac{\Gamma_1^2}{4\Gamma_2} & z >> 1 \\ & \\
                                           \Gamma_1^{2/3}                  & z << 1
                       \end{array} \right.
\end{eqnarray}
In the first case we obtain
\begin{equation}
x \simeq (-\Gamma_2 )^{1/4} - \frac{\Gamma_1}{4 \sqrt{-\Gamma_2}}
\end{equation}
and in the second case
\begin{equation}
x \simeq | \Gamma_1 |^{1/3} \;.
\end{equation}
Using the expressions for $\Gamma_1$ and $\Gamma_2$,  
we obtain the equations 
(\ref{e24})-(\ref{e24b}). 


\end{document}